# Effect of surfactant concentration on the responsiveness of a thermoresponsive copolymer/surfactant mixture with potential application on "Smart" foams formulations.

M. M. Soledad Lencina[1], Eugenio Fernández Miconi[1,2], Marcos D. Fernández Leyes[1], Claudia Domínguez[1,2], Ezequiel Cuenca and Hernán A. Ritacco[(*)1,2].

[1] Instituto de Física del Sur (IFISUR-CONICET), Av. Alem 1253, Bahía Blanca (8000), Argentina

[2] Departamento de Física de la Universidad Nacional del Sur, Av. Alem 1253, Bahía Blanca (8000), Argentina.

(*) Corresponding author: hernan.ritacco@uns.edu.ar

**ABSTRACT**

We studied a system formed by a mixture of a thermoresponsive negatively charged graft copolymer (Alg-g-PNIPAAm) with a brush-type structure, and an oppositely charged surfactant (DTAB), in bulk and at the air-solution interface. We performed experiments of surface tension, electrophoretic mobility, dynamic and static light scattering and atomic force microscopy in order to characterize the complexes formed as a function of DTAB concentration and temperature. We found that these polymer-surfactant complexes are able to respond by changing their sizes, both in bulk and at the air-solution interface, when T is increased above the coil-globule transition temperature (LSCT) of the copolymer. However, the thermoresponse was found to be dependent on surfactant concentration, $c_s$: for $c_s$ < 2.8 mM, the size of the aggregates decreases as T increases but, for $c_s$ ≥ 2.8 mM, the opposite behavior takes place, i.e. the size increases with T. At the interface, the intensity of the effect produced on the surface tension by increasing T above LCST diminishes continuously as $c_s$ increases, reducing the ability of the interfacial complex to respond to temperature changes. We studied the stability of aqueous foams formulated with these mixtures as a function of T and $c_s$. We found that the stability of the foam can be modulated by changing T, but we observed that this effect is dependent on the surfactant concentration range. We found a correlation between changes in the aggregate's sizes, the surface tension behavior and the responsiveness of foam stability to changes of temperature.

# 1. INTRODUCTION

Polyelectrolytes are polymers that dissociate in macroions and small counterions when dissolved in water. Surfactants are generally small molecules whose chemical structure has two distinct parts: the polar head with affinity to polar solvents (water), and the hydrophobic tail with affinity to non-polar fluids. These molecules have the property of adsorbing spontaneously onto the interface separating two immiscible fluids, one polar and one non-polar, such as the air-water interface. At a certain concentration, surfactants self-aggregate in bulk to form micelles[1]. The concentration at which this happens is called Critical Micelle Concentration (*cmc*). Polyelectrolytes and surfactants are used in a broad number of technological applications both on their own and mixed. The richness of the behavior of polyelectrolyte/surfactant mixtures[2] is such that they are used in many industries and are envisaged as systems with great potential for being used in a great number of new technological applications. Among these we find some in the personal care and oil industries, in wastewater treatment, paints, as gene carriers in gene therapy and encapsulation in drug delivery systems, to name but a few [3–6].

We are concerned here with the association between polyelectrolytes and oppositely charged surfactants [2–4,7]. In this case, the association between species is driven by both hydrophobic and electrostatic attraction. The features of the complexes formed and the phase behavior of solutions are the result of an intricate balance between attractive and repulsive interactions among polyelectrolytes and surfactants and depend both on the physical conditions, like pH, temperature or ionic strength, and on the chemical nature of surfactants and polyelectrolytes such as charge density, hydrophobicity of chains, molecular weight, degree of branching, etc., as well as on the concentration of both polyelectrolytes and surfactants. As just stated, the behavior of these systems depends on the specific chemical system[8]; however the following general picture can be given[7]: When an oppositely charged surfactant is added to a polyelectrolyte solution it first progressively replaces the polyelectrolyte counterions in the vicinity of the macromolecular main chain. Generally, this process does not conduct to observable changes in the bulk properties of the system as could be followed with commonly used techniques as conductivity or light scattering, however they can be detected by more sensitive, and less common techniques such as Electric birefringence[9–13]. This situation changes when a certain surfactant concentration, the critical aggregation concentration (*cac*) is reached. At this concentration, surfactant molecules begin to cooperatively bind onto the macromolecule chain. The *cac*, in general, occurs at concentrations 1 to 3 orders of magnitude lower than the *cmc* of pure surfactant solutions, and can be determined by calorimetry, conductivity or surface tension[8] measurements. In this last technique, the *cac* is ascribed to the beginning of the first plateau in the surface tension isotherms[14], this concentration is also known as the T1 point. As surfactant concentration continues to increase, surface tension remains almost constant (plateau) until the T2 point is reached. The T2 concentration corresponds to the saturation of the binding sites onto the polyelectrolyte chain. At this point, surfactant/polyelectrtolyte complexes become hydrophobic and phase separation may occur. At higher surfactant concentrations, in general over the *cmc* of the pure surfactant, redissolution of these precipitates may happen. This last concentration is called T3.

The features of the polymer/surfactant complexes, such as size and shape, have been studied with a number of techniques including dynamic (DLS) and static (SLS) light scattering, X-ray spectroscopy, small angle scattering of X-ray and neutron techniques (SAXS, SANS), among others [7,15–19]. From all these experiments it was found that a great number of factors influence the size and shape of the polyelectrolyte/surfactant complexes as well as the characteristics of the phase diagrams[20,21]. To make things even more complicated, polyelectrolyte-surfactant complexes often remain trapped in non-equilibrium metastable states whose characteristics depend on the history of the systems, for instance on the protocols of mixing or on the time elapsed since preparation[22–26].

Another area of applications of these systems, not mentioned above, is as stabilizing agents in liquid foams (and emulsions). In this respect, properties, such as surface tension and surface rheology, imparted to fluid-fluid interfaces and films due to the presence of polyelectrolyte-surfactant complexes in the regions separating immiscible fluids, is of crucial importance[27–35].

Stabilizing liquid foams was the main motivation of this study, particularly, we desired to formulate liquid foams capable of responding to external stimuli[36]. In this respect it was envisaged the use of a thermo-responsive polymer, PNIPAAm, as foaming agent potententialy capable of responding to temperature changes. PNIPAAm undergoes a conformational transition at about 35°C, being in a coil conformation below this temperature and collapsing to form globules above it. Additionally, it was shown that PNIPAAm can adsorb at interfaces and transit from a fluid-like to a solid-like surface layer when the transition temperature is crossed[37,38]. Because the transition is reversible both in bulk and at the interfaces, PNIPAAm solutions were considered as candidates for the formulation of "smart" foams where their stability could be switched on/off by changing the temperature. Unfortunately, the foaming properties of PNIPAAm aqueous solutions are quite poor and it was found that the foams produced from them were unstable[39], precluding its use as stabilizing agent in foams formulations. Guillermic et al.[39] tried to overcome this problem by mixing the PNIPAAm with the surfactant sodium dodecyl sulfate (SDS) in order to improve the foaming properties of the solutions. Despite they succeeded in this respect, foamability and foam stability were indeed improved, the thermal responsiveness of the interfacial layer was lost.

With the intention of producing a foaming system capable of responding to changes in temperature, we synthetized a copolymer based on PNIPAAm and alginate, which is a negatively charged polysaccharide (see supporting information, ESI), to give place to a negatively charged polyelectrolyte with a brush-type structure, capable of forming complexes with oppositely charged surfactants. Because the PNIPAAm are incorporated as side chains, we speculated that, when mixed with an oppositely charged surfactant molecule, the responsiveness to temperature changes of the system would be maintained and that, at the same time, the foamability properties and the stability of foams would improve.

In this article we study a mixture of the copolymer Alg-g-PNIPAAm, hereafter called Cop-L, with the cationic surfactant dodecyltrimethylammonium bromide (DTAB), which was characterized by surface tension, electrophoretic mobility and zeta($\zeta$)-potential, dynamic and static light scattering (DLS, SLS) and atomic force microscopy (AFM) as a function of surfactant concentration and temperature. Our goal was to study the structure of aggregates in bulk and its dependence on temperature, in order to evaluate their potential use and performance as foam stabilizers in the formulation of responsive ("smart") foams.

In the present article we focus mainly on the properties of the polyelectrolyte/surfactant aggregates in bulk, in a subsequent work we will systematically explore the interface properties including dynamic surface tension and interfacial rheology and its coupling with the behavior of foams (foamability and foam stability) formulated with them. We only present in this article some preliminary results on foam stability below and above the polyelectrolyte transition temperature, which show that the system is in fact capable of responding to temperature changes.

## 2. MATERIALS AND METHODS

### 2.1 Materials

The cationic surfactant, dodecyltrimethylammonium bromide (DTAB) was obtained from Sigma-Aldrich (99%) and used as received.

Sodium alginate (Mw=198) is the sodium salt of alginic acid, a lineal polysaccharide obtained from brown algae constituted by two uronic acids as repetitive units, 1,4 b-D-mannuronic acid (M) and 1,4 a-L-guluronic acid (G), in the form of homopolymeric (MM- or GG-blocks) and heteropolymeric sequences (MG- or GM-blocks). A low viscosity sodium alginate was purchased from Alfa Aesar with a mannuronic/guluronic ratio (M/G) estimated to be 2.2 by $^1$H NMR according to the literature[40–42]. Poly(N-isopropylacrylamide, PNIPAAm, is a synthetic polymer that presents a low critical solution temperature (LCST) undergoing a volume phase transition when heated. At low temperatures, intermolecular hydrogen bonds between water and polar groups of PNIPAAm solubilise the polymer. Above the LCST hydrogen bonds break and hydrophobic associations between polymer chains take place, resulting in a collapsed state. The LCST for high molar mass PNIPAAm is around 32°C, but this critical transition temperature is a function of molar mass and polymer concentration, among other parameters[38,43–46].

The alginate-g-PNIPAAm graft copolymer (Cop-L) was obtained by a coupling reaction between the carboxyl groups of sodium alginate and the terminal amine groups of PNIPAAm-$NH_2$ chains, using 1-ethyl-3-(3´-(dimethylamino) propyl) carbodiimide hydrochloride (EDC) as the coupling agent. Thus, a brush-type anionic polyelectrolyte was synthesized with Mn= 4200 g/mol PNIPAAm side chains. The synthesis and further characterization were extensively described in a previous work[47]. The mean molecular weight of the co-polymer was determined by static light scattering giving a value of Mw = 89.5 KDa. The number of charges per co-polymer molecule was found to be about 300.

Polyelectrolyte solutions were prepared by dissolution in ultrapure water (Milli-Q water purification system). Due to the limited amount of polymer available, a unique and fixed polymer concentration ($c_p$) of 400 mg.L$^{-1}$ was used in the preparation of all samples, except for electrophoretic mobility experiments in which, in order to obtain the binding isotherms (see next sections), we also used solutions of $c_p$= 100 mg.L$^{-1}$.

### 2.2 Sample Preparation protocols and measurements.

Two different protocols of sample preparation were used. For surface tension measurements, a concentration process was employed. First, the surface tension of a DTAB free aqueous solution of Cop-L at $c_p$= 400 mg.L$^{-1}$ was measured. Subsequently, proper amounts of the copolymer and DTAB solutions, and water were added until the targeted concentration was achieved, the surface tension was then measured after an equilibration period of not less than 60 minutes. This process was repeated until the whole range of DTAB concentration was covered.

For DLS, SLS, mobility and ζ-potential measurements, all samples were obtained by adding equal volumes of the DTAB solution with double the desired final concentration to 800 mg.L$^{-1}$ of the Cop-L solution. Solutions were left to reach equilibrium for 24h prior to measurement. Some of these bulk experiments, DLS and ζ-potential, were repeated with samples prepared following the first protocol of preparation and we found no significant differences in the corresponding results.

### 2.3 Methods

#### 2.3.1 Surface tension

Surface tension measurements were carried out using the sensor of a Langmuir balance (KSV NIMA) and a paper Wilhelmy plate. Experiments at room temperature were performed using a Teflon trough (10 ml of volume) while a jacketed vessel was employed for temperature-dependent measurements.

Pure water surface tension measurements were used to verify optimal paper probe quality before each experimental iteration. After solutions were poured into the corresponding vessel, surface tension was continuously measured until a stable value was achieved. The reproducibility was ± 0.2 mN m$^{-1}$.

Temperature dependent experiments were performed in the range of 20 to 55 °C, with measurements being taken every 5 °C. An approximated heating rate of 1 °C/min was used between steps. Once the required temperature was reached, samples were left to reach equilibrium for 30 to 60 minutes before surface tension determination. Temperature was controlled using an external circulating water bath (Lauda Alpha) and, it was monitored by means of a thermocouple.

### 2.3.2 Dynamic (DLS) and Static (SLS) Light Scattering

The aggregate sizes of Cop-L/DTAB complexes were measured as a function of temperature and DTAB concentration by DLS. A Malvern Autosizer 4700 with a Series 7032 Multi-8 correlator and equipped with 20 mW laser (OBIS Coherent) operating at a wavelength ($\lambda$) of 514 nm were employed, with detection at scattering angles ($\theta$) between 30 and 150°. The intensity auto-correlation functions were processed by the Autosizer 4700 software using Cumulants or CONTIN analysis and the apparent translational diffusion coefficients, $D_{app}$, obtained for each scattering angle. The mean translational diffusion coefficients, $D_s$, were obtained by extrapolating $D_{app}$ to $q^2=0$, being q the wave vector (q= 4$\pi$ n sin($\theta$/2)/$\lambda$, where n is the solvent refractive index),

$$D_{app}(q) = \langle D_s \rangle (1 + Kq^2) \tag{1}$$

Once $D_s$ was obtained, the hydrodynamic radius, $R_H$, was determined from the Stokes-Einstein equation,

$$D_s = \frac{k_B T}{6\pi \eta R_H} \tag{2}$$

Being $k_B$ the Boltzmann constant, T the temperature and $\eta$ the solvent viscosity. The temperature was controlled (± 0.1 °C) using the device´s own system (PCS 8 Temperature Controller) and an external circulating water bath (Lauda Alpha).

The intensity of the light scattered by the samples was measured with the same device as a function of q at angles between 20 and 150°, by steps of 1°. The corresponding dependence of the scattered intensity on q, I(q) (and form factors, P(q)~I(q)), were analyzed by means of the Guinier-Porod empirical law [48–50],

$$I(q) \sim \frac{1}{q^s} \exp\left[-\frac{q^2 R_g^2}{3-s}\right] \quad for\ q \leq q_l = \frac{1}{R_g}\sqrt{\frac{(m-s)(3-s)}{2}}$$

$$I(q) \sim \frac{1}{q^m} \quad for\ q \geq q_l \tag{3}$$

With *m* being the Porod exponent, $R_g$ the radius of gyration and *s* a dimensional variable (for 3D globular objects, such as spheres, s = 0; for 2D symmetry, such as for rods, s = 1 and for 1D objects, such as for lamellae or platelets, s = 2). When applicable, the form factor was fitted with the model for homogeneous spherical particles,

$$P(q) = \frac{9}{(qR)^6}[sin(qR) - qRcos(qR)]^2 \qquad (4)$$

With R being the radius of the sphere. In the fitting procedure we used a smearing function for a pinhole (Gaussian), which was determined for our device by measuring the form factor for a Latex particle standard of 500 nm.

### 2.3.3 Electrophoretic Mobility and ζ-Potential

The electrophoretic mobility and zeta(ζ)-potential[51] of polyelectrolyte/surfactants aggregates were measured using a Malvern Zetasizer Nano ZSP from Malvern Instruments (light source 10 mW He-Ne laser, wavelength 633 nm). This instrument uses the laser Doppler velocimetry method with Phase Analysis Light Scattering (PALS) in order to obtain the electrophoretic velocity, v, of colloidal particles and from it the mobilities, u=v/E, being E the electric field applied. Once u is measured, the ζ-potential is calculated by means of the Henry equation and Smoluchowsky approximation, $\zeta = \eta\,(u/\epsilon)$, with η and ε the solvent viscosity and permittivity respectively.

Each mobility value obtained is an average of several measurements, according to Malvern´s proprietary "Quality Factor" statistical criterion.

Disposable capillary cells were used. Samples were allowed to reach their equilibrium temperature for 60 minutes prior to experiments. Values were taken in triplicate with a delay of 120 seconds in between.

### 2.3.4 AFM

Atomic force microscopy (Bruker Innova) measurements were performed under ambient conditions in tapping mode using RTESP-CP tips (Veeco, spring constant = 20-80 N/m, as reported by manufacturer). Samples were prepared by casting drops of solutions containing copolymer-surfactant mixtures onto a smooth glass surface, and then evaporating the water in a vacuum chamber. Images with a scan range of 3 µm at a scan rate of 1 Hz were taken and processed using the Gwyddion software.

### 2.3.5 Viscosity measurements.

The intrinsic viscosities of selected complexes solutions with DTAB concentration 0; 0.3; 1.62; 2.82 and 15 mM were determined at 25 and 45 °C using an Ubbelohde viscometer. The values reported are the average of 10 measurements.

### 2.3.6 Experiments on Foams.

In order to evaluate the properties of foams formulated with the Cop-L/DTAB mixtures, we produced foams by means of two syringes connected through a tube of very small internal diameter (Tygon internal diameter = 1/16 inch, length 10 cm) as explained in the literature[52,53]. One of the syringes was filled with the desired volumes of air, $V_g$, and foaming solution, $V_l$, in order to fix the initial liquid fraction of the foam, $\phi_{l,0} = V_l/V_{foam}= V_l/(V_l+V_g)$. The liquid and air were then transferred from one syringe to the other through the constriction given by the small cross section tube, in a series of 10 cycles. In all the experiments presented in this article $\phi_{l,0}$ was fixed to 0.25. Bubbles produced by this device had a mean radius of 70 µm. The foam so produced was then transferred to a rectangular glass cell (Hellma, OS) with a light path of 1 cm which was placed into a homemade holder adapted to a UV-vis spectrometer of fiber–optic (Ocean optics USB2000+) as shown in figure 1. Solutions and cells were thermalized prior to foam production. A CCD camera (Basler, acA1300-30um) was placed in front of the cell. The light emitted by a Xenon lamp (Ocean Optics PX-2) was sent through the foam sample via a fiber-optic placed at half of the cell's height and the transmitted intensity was collected by a second fiber-optic and measured with the UV-vis spectrometer (by integrating the whole spectrum) as a function of time, (one spectrum per second was taken and saved in a computer for analysis). With this setup we simultaneously followed the foam height, the volume of liquid drained and the transmitted light intensity as a function of time (see fig. 1).

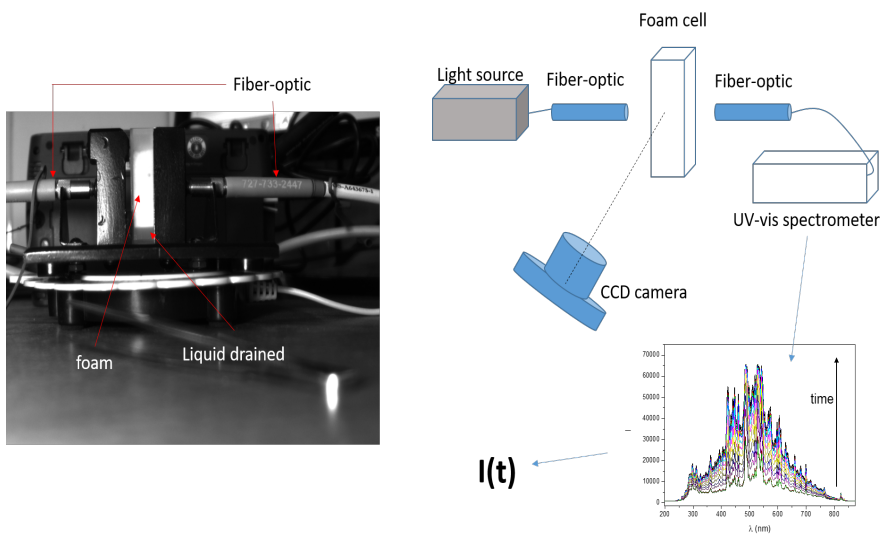

Figure 1: Scheme of the device used to study the foams formulated with Cop_L/DTAB mixtures.

## 3. RESULTS.

### 3.1. Equilibrium Surface tension isotherms.

Surface tension measurements were carried out on several aqueous solutions with increasing DTAB concentration ($c_s$) and a fixed Cop-L concentration, $c_p$= 400 mg.L$^{-1}$. Measurements were performed at two temperatures, 25 °C and 45 °C. The results are shown in Figure 2. First, it is important to note the significant drop in surface tension caused only by the alginate-g-PNIPAAm copolymer (Cop-L), displaying a clear surface activity. The surface pressure, $\Pi = \gamma_0 - \gamma$ (being $\gamma_0$ and $\gamma$ the surface tension of pure water and solutions respectively), was 26.6 mN.m$^{-1}$ and 30.2 mN.m$^{-1}$, at 25°C and 45°C, respectively.

Regarding the effect of DTAB on surface tension, figure 2 shows the presence of two plateaus. For the measurements at T=25°C, the first plateau begins at a surfactant concentration of about $c_s$~ 0.7 mM (T1 on the figure) and ends at about $c_s$~ 7 mM (T2 on the figure). Then, as $c_s$ increases, the surface tension drops until the second plateau begins at T3, which is close to $c_s$ ~ 16 mM. From then on, the surface tension remains constant up to the highest surfactant concentration used, $c_s$ ~ 80 mM. A similar behavior is observed for T= 45°C, in this case T1 is about 0.5 mM while T2 and T3 occur at the same concentrations.

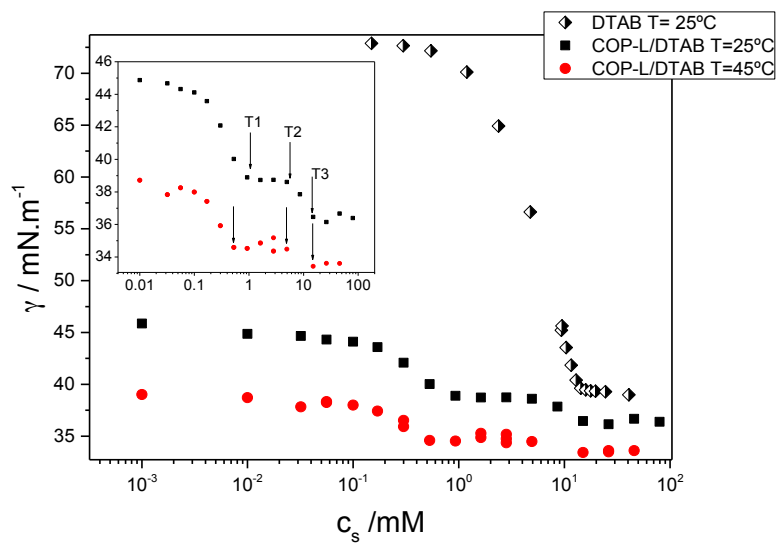

Figure 2: : Surface tension of Cop-L/surfactant mixtures as a function of DTAB concentration at 25°C and 45°C.

## 3.2 Surface tension as a function of temperature.

Figure 3 presents the behavior of surface tension as a function of temperature for different DTAB concentrations. Surface tension values at a fixed temperature decreased with increasing DTAB concentration, as expected. For constant $c_s$, all solutions studied showed a linear decrease with temperature, interrupted by a notable change in slope. The intersections between lines of different slopes were found to be around 39-43 °C in all cases. These results are related to the presence of the low critical solution temperature (LCST) of PNIPAAm moieties.

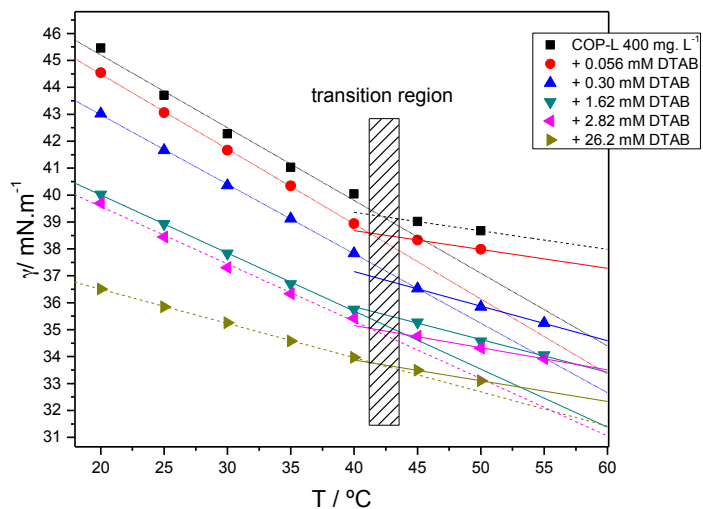

Figure 3: Surface tension as a function of temperature for several Cop-L/DTAB mixtures. A transition at T>LCST is clearly seen.

### 3.3 Phase Behavior.

The phase behaviour of mixed Cop-L/DTAB solutions was observed as a function of temperature and surfactant concentration. In figure ESI-2 in the supporting information (ESI) the aspect of solutions for four different surfactant concentrations at a temperature of 20 °C is shown. At this temperature and for all surfactant concentrations from 0 to 30 mM the suspensions are stable and no phase separation was observed. As the temperature increases from 20 to 55 °C, phase separation is observed for surfactant concentrations between 8 and 15 mM. Below and above this concentration range the systems are stable (no precipitate) at all temperatures.

### 3.4 Dynamic (DLS) and Static Light Scattering(SLS): size and geometry of aggregates.

In order to obtain information on the size of the aggregates we performed DLS experiments. We measured the hydrodynamic radius, $R_H$, at four scattering angles, $\theta$= 30, 60, 90 and 120 degrees. First, values of $R_H$, as a function of temperature, for the polyelectrolyte alone were obtained. A sharp transition temperature, LCST, of 38±1°C, with $R_H$ going from about 1000nm, below the LCST, to 350 nm, above it, was found. In these samples the correlation functions were well fitted with monoexponentials (at least in the time range explored, see ESI) and the characteristic times were found to depend slightly on the scattering angle. Figure 4 presents the hydrodynamic radius, $R_H$ as a function of DTAB concentration, for both temperatures, above and below the transition temperature.

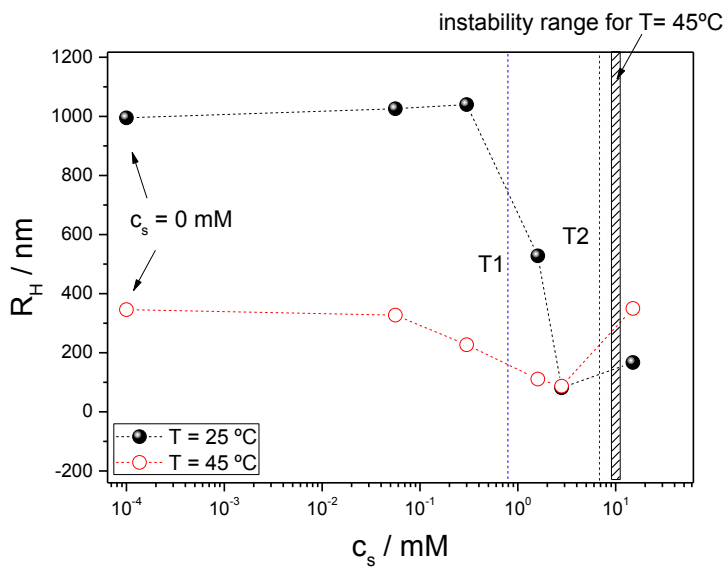

Figure 4: Hydrodynamic radius of polymer/surfactant complexes in aqueous solution, $c_p$= 400 mg.L$^{-1}$, as a function of DTAB concentration at 25°C (closed circles) and at 45°C (open circles). The points corresponding to Cop-L 400 mg.L$^{-1}$ without surfactant were included out of scale ($c_s$ = 0).

In the figure we observe that $R_H$ diminishes by a factor of about 4 as T becomes higher than the LSCT for all mixtures with $c_s$<0.5mM. For 0.5 < $c_s$ <2.8 mM the change in $R_H$ when T crosses the transition temperature, diminishes continuously, and becomes very small at a surfactant concentration of 2.8 mM (see also fig. 7b). For concentrations higher or equal to 2.8 mM, the opposite is true, $R_H$ increases as the temperature passes from 25 to 45 °C. We also observe that the collapse produced by the addition of surfactant at a concentration over 1 mM, at the lower temperature, is equivalent to the collapse produced on the free surfactant polymer solution by changing the temperature above LCST. The polymer collapse at $c_s$ ~1 mM is also observed by viscosity measurements (see ESI).

The polydispersity index (PI) obtained from cumulants analysis of the intensity auto-correlation functions, are between 0.3 and 0.05 for all samples with $c_s$> 1 mM. For free

DTAB Cop-L solutions and mixtures with $c_s$<1.6 mM, both at T= 25°C, the obtained PI were between 0.5 and 1, in those cases we used CONTIN analysis. For the same solutions but at T> LCST, the PI were below 0.3. These results indicate quite monodisperse aggregates both when $c_s$ > 1.6 at low temperature, and for T> LCST.

In order to gain information on the form of these aggregates we performed measurements of the intensity of the scattered light as a function of the scattering angle for some of the samples. The concentrations studied were those corresponding to $c_s$= 0 (pure Col-L); $c_s$= 1.6 and $c_s$= 2.8 mM. In figure 5 we show the intensity of light scattered as a function of wave vector q (form factor), for a mixture with $c_p$= 400 mg. L$^{-1}$ and $c_s$= 1.6 Mm, at T= 25°C. Oscillations in the scattering intensity function are clearly seen in this case. The line in the figure corresponds to the fitting curve obtained with eq. 4 (with pinhole smearing), which gave R= 451 nm. The experimental points shown in figure 5 are the result of averaging 3 independent measurements, each of which was measured 6 times. To make sure that the oscillations were not a consequence of imperfections on the walls of the cylindrical cells, we rotated the sample cell by 60° in between measurements, until a complete turn of the cell was completed. Note that the value R= 451 nm is close to the value obtained by DLS, $R_H$= 528 nm.

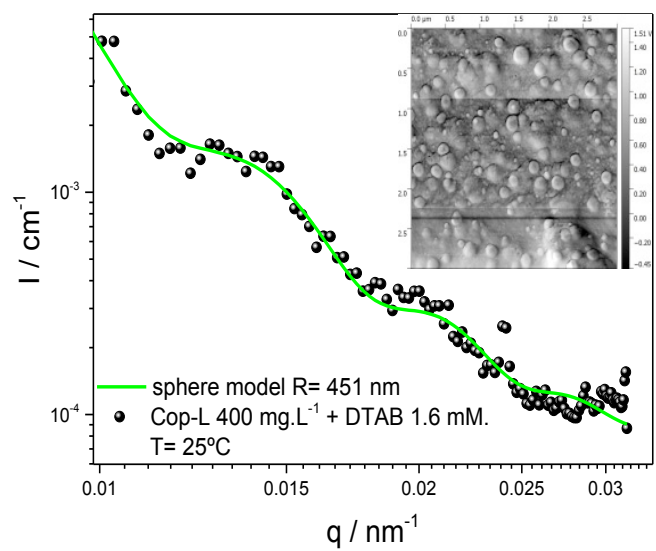

Figure 5: Experimental Form factor (circles) for the systems Cop-L 400 mg.L$^{-1}$ + DTAB 1.6 mM at 25°C. The line corresponds to the fitting with a sphere model (see text). The inset is a AFM image obtained for a mixture of Cop-L 400 ppm/DTAB 1.6 mM deposited onto a Si-wafer and dried.

In figure 6 we show the form factor for a Cop-L solution with $c_p$= 400 mg L$^{-1}$ without surfactant, at T= 45°C (fig. 6a) and for a mixture with $c_p$= 400 mg L$^{-1}$ and $c_s$= 2.8 Mm, at T= 25°C (fig. 6b). In the figures we include fittings with the Guinier-Porod empirical law (eq.3). From the fittings we found Rg= 348 nm for the former solution. Because multiple scattering is present and the Rayleigh-Debye-Gans limits are not fulfilled in this case (qRg<1), the results should be taken with a pinch of salt, however, the ρ-ratio, ρ=$R_H$/Rg = 345/348 ~1, is consistent with spheroidal aggregates. For the solution with 2.8 mM of DTAB we found from the fittings with eq. (3), Rg= 167 nm, ρ=$R_H$/Rg = 125/167 ~0.75, which is what one would expect for spherical aggregates. The values found for the Porod exponent, m, and for the dimensional parameter s, were compatible with globular aggregates with fractal surfaces (m~2.5, s~0.5), for both solutions.

In order to confirm these results we performed atomic force microscopy (AFM) experiments for the mixtures with $c_s$=1.6 and $c_s$= 2.8 mM. In the inset of figure 5, an AFM image of those aggregates is shown, both the size and form of aggregates are compatible with DLS and SLS results (see also ESI).

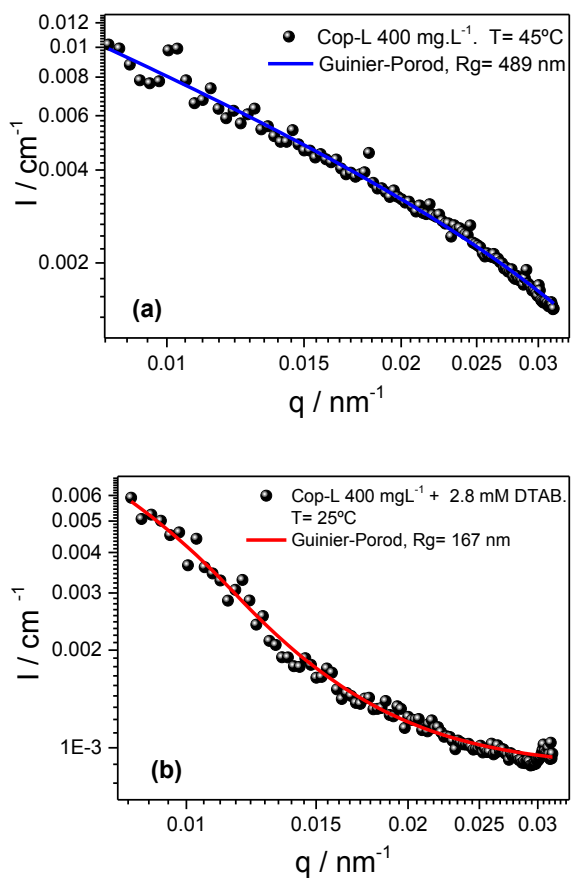

Figure 6: Experimental form factor (circles) for the systems (a) Cop-L 400 mg.L$^{-1}$ at 45°C; (b) Cop-L 400 mg.L$^{-1}$ + 2.8 mM DTAB. The lines correspond to the fitting with the Guinier-Porod empirical model.

For each surfactant concentration, the hydrodynamic radius as a function of temperature was also measured by DLS. In figure 7 we show the change in $R_H$ as the temperature increases for Cop-L solution at $c_p$= 400 mg.L$^{-1}$ and for the mixed system with 2.8 mM of DTAB. In the first case (fig. 7a) $R_H$ diminishes abruptly from about 1300 to 300 nm when the transition temperature is crossed. This corresponds to the system with the maximum change in size. The minimum variation of $R_H$ is found for the system with 2.8 mM of DTAB (fig. 7b). For all systems with higher surfactant concentrations the hydrodynamic radius increases as T becomes higher than the transition temperature.

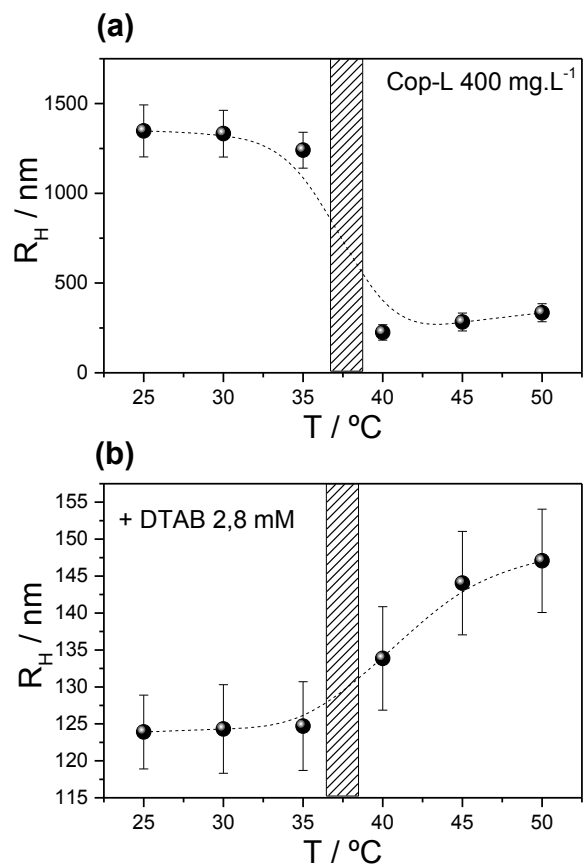

Figure 7: Hydrodynamic radius as a function of temperature measured by DLS (a) Cop-L 400 mg.L$^{-1}$ solutions; (b) Cop-L 400 mg.L$^{-1}$ + DTAB 2.8 mM.

### 3.6 Electrophoretic Mobility measurements.

In figure 8a we present results of electrophoretic mobility for the mixed polyelectrolytes/surfactant system for two different polymer concentrations and at a temperature of 25°C. In figure 8b, we show the ζ-potential as a function of surfactant concentration for two temperatures, above and below the LCST. The curve for T= 45°C is qualitatively similar to that at 25°C, except for the small region were a precipitate appears, indicated by a bar in said figure.

Note that the mobility and ζ-potential become zero at a total surfactant concentration of about 15 mM ($c_p$=400 mg.L$^{-1}$), which coincides with the surfactant *cmc*. It has been shown that, under certain conditions, the amount of surfactant molecules bound to the polyelectrolyte can be estimated from measurements of electrophoretic mobility[54] for two different polymer concentrations, we will use results on figure 8a with that purpose. (see discussion below).

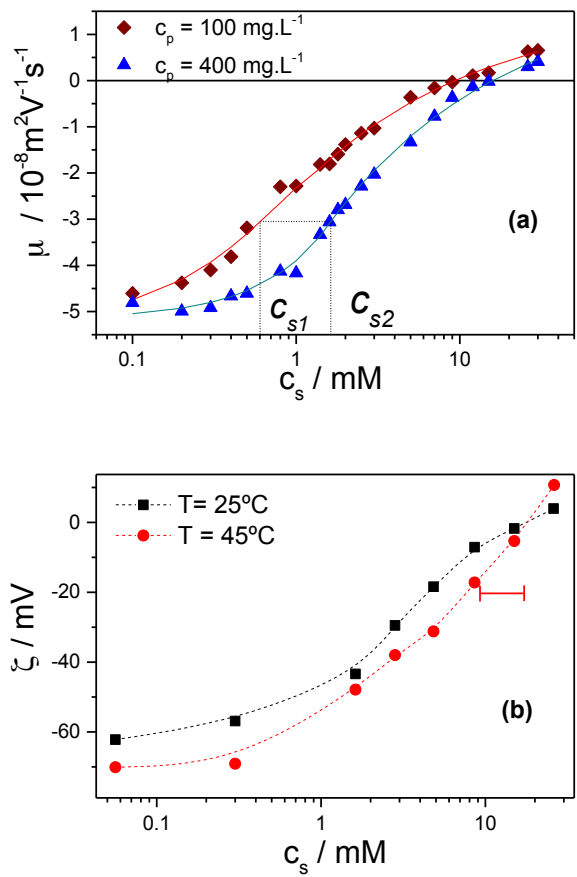

Figure 8: (a) Electrophoretic mobility of Alg-PNIPAM/DTAB complexes versus DTAB concentration. Solid lines correspond to ad hoc fitted functions, used to interpolate u values. Dotted lines indicate schematically how DTAB concentration belonging to equal values of u are determined (see discussion). (b) Zeta potential from mobility measurements as a function of DTAB concentration and at two temperatures, below and above the LCST.

## 4. DISCUSSION.

Phase behavior and Surface tension.

The surface activity shown in figure 2 by the surfactant free solutions, $c_s = 0$, is mainly attributed to the presence of PNIPAAm on the copolymer chain, since alginate aqueous solutions do not show significant surface activity at similar concentrations[56]. In this sense, Zhang et al.[57] reported a sizable effect of PNIPAAm on the surface tension of aqueous solutions even for concentrations as low as 5 mg.L$^{-1}$.

From the surface tension isotherms (figure 2) we identified three characteristic surfactant concentrations: T1, T2 and T3. The concentration T1, which corresponds to the beginning of the first plateau, is generally associated to the Critical Aggregation Concentration, *cac*, and corresponds to the onset of binding of DTAB to alginate-g-PNIPAAm in bulk. Upon further increase of the amount of surfactant, the polymer saturation point (T2) is reached. At this point, it is assumed that all of the binding sites of the polymer are occupied by surfactant molecules and any excess causes a decrease in surface tension until the critical micelle concentration (*cmc*) is reached. Note that the concentration T2 (~7mM) is below the concentration at which the electrophoretic mobility approaches zero (~ 15 mM, see fig. 8a), which is close to T3. Above T3, any DTAB addition would lead to the formation of micelles probably decorated with polymer chains, with no effect on surface tension[58]. Besides the overall decrease in surface tension previously mentioned, temperature increment seems to cause a slight shift of T1, probably due to an increased hydrophobicity interaction between polymer and surfactant. Also, in contrast to the behaviour observed at 25 °C, at 45 °C the polymer precipitated in a concentration region between 8 and 15 mM, i.e. between T2 and T3, this is also attributed to the increased hydrophobicity of the aggregates at the higher temperature. At concentrations above T3, the precipitates are redissolved, leading to stable dispersions. This last concentration coincides with the *cmc* of the surfactant and also with the surfactant concentration region where a size increment is observed as T increases over the transition temperature (see DLS data), thus we interpret this as indication of a change in the structure of the aggregates.

The effect of DTAB and temperature at the interface is more clearly seen in figure 3. For pure liquids, the slopes of surface tension vs. temperature curves are related to the surface entropy, $S^s = -\frac{\partial \gamma}{\partial T}$, therefore, the changes in the slopes, m ($m = \frac{\partial \gamma}{\partial T}$, from figure 3) can be related to changes in the surface entropy. The relative changes in the slopes, $m_r$ when the transition temperature is crossed are shown in figure 9 as a function of DTAB concentration. The relative slope change, $m_r$ is defined as,

$$m_r = \frac{m_{T<LCST} - m_{T>LCST}}{m_{T<LCST}} \tag{5}$$

Where $m_{T<>LCST}$ stands for the slopes below (<) and above (>) the LCST. For free surfactant polyelectrolyte solutions, the reduction of the slope is about 75%, suggesting an entropy reduction as T becomes higher than the transition temperature. This can be rationalized in terms of the conformational changes occurring on the polymer chains, which go from coil to globule, at the interface. As the DTAB concentration increases, it induces a progressive collapse of the polyelectrolyte at temperatures below LCST (see figure 3) then, the conformational changes observed when crossing the transition temperature are less and less pronounced. This is what we observe from the relative changes in the slopes in figure 9.

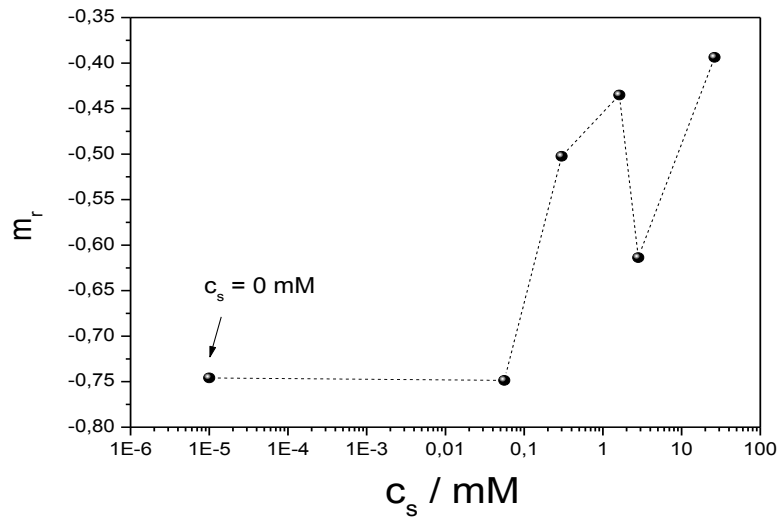

Figure 9: Relative change of slopes of surface tension vs. T curves when crossing the transition temperature at each surfactant concentration as obtained from figure 2.

DLS and SLS.

The results of DLS are similar to those found for DTAB/CarboxyMC (sodium corboxymethylcellulose)[59]. The addition of an oppositely charged surfactant to a flexible polyelectrolyte produces, at certain concentrations, the polymer collapse. This results in aggregates which are spherical and monodisperse, as evidenced by DLS and SLS (figs. 4 and 5) results. As stated in reference[59] the monodispersity of the aggregates is quite surprising, if ones takes into account that the size distribution of the polyelectrolyte chain is rather broad.

Figure 4 clearly shows that the addition of DTAB produces, at $c_s$= 2.8 mM, a hydrophobic collapse of the polymer chain in a way similar to that produced by an increment of T above LCST for Cop-L solutions without DTAB. This seems to indicate that the DTAB molecules bind mainly to the PNIPAAm side chains instead of the charged groups on the alginate

(see also ζ-potential results). This collapse of the polymer chain as DTAB concentration increases is also observed by viscosity measurements (see ESI).

The effect of the aggregate size increasing after the collapse, as DTAB concentration increases, was also observed in the DTAB/carboxilMC. This suggests a change in the structure of the aggregates as $c_s$ increases above T2 (fig. 4).

### Electrophoretic mobility, ζ-potential and binding isotherms.

According to Mezei et al.[54] the binding isotherms of ionic surfactants on oppositely charged polymers can be estimated from electrophoretic mobility data of surfactant/polymer complexes. The relative amount of surfactants bound to polymer, B can be expressed as

$$B = \frac{c_s - c_f}{c_p} \quad (6)$$

Where $c_s$ is the total surfactant concentration, $c_f$ is the equilibrium free surfactant concentration, and $c_p$ is the polymer concentration. Equation (6) is valid if a $c_f$ is smaller than the *cmc* and if either the ionic strength of the solution is high, or the charge density and the polymer concentration is not too high.[54,60].

In order to calculate B using eq. 6, it is assumed that B depends only on $c_f$, and that the electrophoretic mobility *u* is a function of B and is independent of the polymer concentration. For two polymer concentrations $c_{p1}$ and $c_{p2}$, the same $B(c_f)$ and thus electrophoretic mobility *u*, will be reached at two different surfactant concentrations $c_{s1}$ and $c_{s2}$.

$$B(c_f) = \frac{c_{s1} - c_f}{c_{p1}} = \frac{c_{s2} - c_f}{c_{p2}} \quad (7)$$

$$u(c_{p1}, c_{s1}, c_f) = u(c_{p2}, c_{s2}, c_f) \quad (8)$$

The determination of $c_f$ is possible by means of eq. 7 using interpolated values of $c_s$ which correspond to equal mobilities, according to eq. 8.

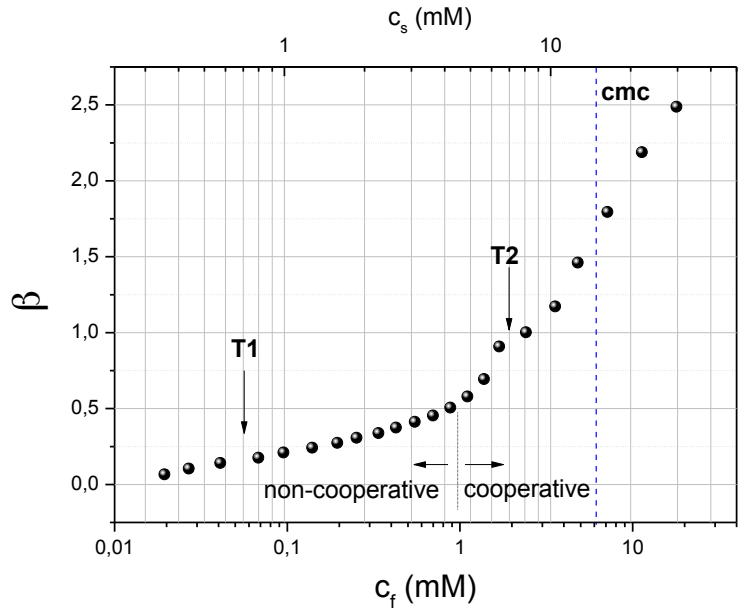

Figure 10: Binding isotherm of DTAB, β is the number of DTAB molecules bound divided by the number of binding sites on the polyelectrolyte chain and $c_s$ and $c_f$ are the total and free (not bound) DTAB concentrations, respectively. The concentrations T1 and T2 obtained from surface tension isotherms are marked with arrows.

In figure 8a we show the mobilities measured at two different polymer concentrations and in figure 10 the corresponding binding isotherm calculated from them using the procedure just outlined. From the mobilities we obtained the free surfactant concentration but, in order to calculate the degree of binding, β, defined as β=(bound surfactant)/(binding sites on the polyeletrolyte), it is necessary to know the number of binding sites on the polymer chain. In general, for polyelectrolyte/oppositely charged surfactant mixtures, the number of charged groups on the polymer chain is taken as the number of binding sites. However, this assumption is not necessarily correct, especially if other kind of interactions, a part from electrostatic, are involved. Thus, from the surface tension isotherm shown in figure 2 and if we, as usual, interpret the T2 concentration as indicating the saturation point in the polyelectrolyte/surfactant association process, that concentration should correspond to the total number of binding sites on our copolymer. The total surfactant concentration at T2 is $c_s(T2)$ = 7 mM, the free surfactant concentration obtained from mobilities and equations 6 to 8 at T2 is $c_f(T2)$= 2 mM (see top and bottom scales in figure 10). The concentration of bound surfactants is then $c_b(T2)$= $c_s(T2)$-$c_f(T2)$ = 5mM. In order to calculate β in figure 10, we used the concentration $c_b(T2)$, as the concentration of binding sites on the polyelectrolyte, instead of the number of charged groups.

The concentration of charged monomers at $c_p$= 400 mg. L$^{-1}$ is 1.38 mM, then, at T2 we have approximately 4 DTAB molecules per charged group associated to the polymer chain. However, note that the $\zeta$-potential becomes zero at a total surfactant concentration of about 15 mM (figure 8b, T=25°C), which corresponds to the surfactant *cmc* and to $c_b$ = 7 mM. From this, it seems that the association process is driven mainly by hydrophobic interactions among the polymer and surfactant molecules, probably involving the PNiPAAM side chains of the brush copolymer. Because the charge inversion occurs at concentrations of bound surfactants 7 or 6 times larger than the number of charged groups on the copolymer, some of the DTAB counterions (Br$^-$) must be condensed onto (into) the polymer/surfactant aggregates. Note that the charge inversion occurs at total surfactant concentrations over the *cmc* of the surfactant (fig. 8). This could indicate the presence of micelles decorating the polymer/surfactant aggregates which would produce an increment of the aggregate's size, which is consistent with the observed increment on the hydrodynamic radius measured by DLS (fig. 4) over the *cmc*. This picture is quite different from what was found for other homopolyelectrolyte/surfactant[4,55] and polyelectrolyte-copolymer/surfactant mixtures[61–63], where the structure of the aggregates is compatible with surfactant micelles decorated with polymer chains, bound together via electrostatic interactions between the charged micelles and the oppositely charged groups on the polyelectrolyte chains.

Returning to the binding isotherm of figure 10, note that the slope of the $\beta$ vs $c_f$ curve indicates a non-cooperative association process up to $c_f \sim 1$ mM where the binding process becomes more cooperative. Close to T2 an inflexion point seems to be present suggesting the typical sigmoidal shape of the polyelectrolyte/surfactant binding isotherms ($\beta$=1 at the binding saturation point). At concentrations above T2, the amount of surfactant molecules bound to the polymer chain increases sharply, which would indicate, as stated before, the presence of a few micelles decorating the aggregates. We recall that the method used to obtain the isotherms is valid for $c_s$<*cmc*, the values for surfactant concentrations above the *cmc* should be considered cautiously.

**Effect on Foam stability. Preliminary Results.**

Our original interest on this complex polymer/surfactant system was because of the possibility of using it to produce thermoresponsive foams. In light of figure 3 we chose to study foams stabilized with solutions at a fixed polymer concentration of 400 mg.L$^{-1}$ and mixed with DTAB at surfactant concentrations of 0.3; 1.5; 2.8 and 20 mM, in an attempt to find a correlation between foam stability and structural changes. Recall that at 0.3 and 1.6 mM there is a reduction in the size of the aggregates (see figure 3) when T goes over the transition temperature, on the other hand, for $c_s$= 2.8 mM and $c_s$=20 mM there is an increment in the aggregate's sizes (see figure 4 and 7) when T crosses the LCST.

In figure 11 a plot of the relative light intensity transmitted through the foam samples as a function of time for four DTAB concentrations, is shown. The relative intensity is defined as: I-I$_0$/I$_{max}$, being I, I$_0$ and I$_{max}$ the instantaneous, I(t), initial, I(t=0) and final (without foam) transmitted light intensities respectively. Because the fiber optic is placed at the middle of the foam container, the time at which the relative intensity reaches a value of 1 indicates the moment when the foam sample (foam + liquid drained) has half its initial height, and the corresponding time, $t_{1/2}$, indicated by arrows on figure 11, is a measure of foam stability. Figure 11a corresponds to a solution of $c_s$= 20 mM without polymer at two temperatures, 20 and 45 °C. This result is used for comparative purposes. Figure 11b and 11c show results for two mixtures of Cop-L and DTAB at two surfactant concentrations, $c_s$= 1.62 mM and $c_s$= 2.82 mM. In table 1 we present all results for $t_{1/2}$ at both temperatures.

| [DTAB]/mM | $t_{1/2}$ (20°C) | $t_{1/2}$ (45°C) | $t_{1/2}$(20°C)/$t_{1/2}$(45°C) |
|---|---|---|---|
| 20* | 500 | 195 | ~2.6 |
| 20 | 1850 | 100 | ~18 |
| 2.82 | 1000 | 180 | ~5 |
| 1.62 | 2000 | 200 | ~10 |
| 0.3 | 3250 | 300 | ~11 |

*Table 1: Foam stability measured by the time needed to reach half the initial foam height, $t_{1/2}$. 20\* is for free polymer surfactant solutions. The polymer concentration for all measurements was $c_p$= 400 mg.L$^{-1}$. The time is given in seconds.*

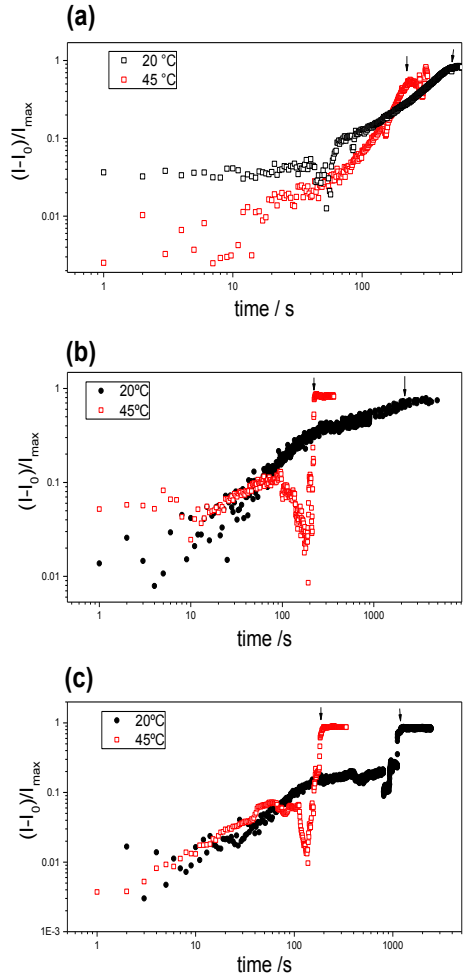

Figure 11: Light intensity as a function of time. (a) DTAB 20 mM; (b) Cop-L 400 mgL$^{-1}$+DTAB 1.62 mM; (c) Cop-L 400 mg.L$^{-1}$ + DTAB 2.82 mM.

Note that for the mixture of Cop-L/DTAB at $c_s$= 2.82 mM the foam is about 5 times more stable at T= 20°C than at T= 45°C. This is similar to the behaviour of DTAB solutions with $c_s$= 20 mM free of polymer (see table, 20*). However, for the mixed systems with DTAB concentrations of $c_s$= 1.62 , $c_s$= 0.3 and $c_s$ = 20 mM, the foam stability at low temperature is between 10 and 18 times larger. These results seem to correlate well with the behaviour observed on the aggregate's size in bulk when T changes from 20 to 45°C, see fig. 4. It is worth noting that foams cannot be stabilized by Alg-PNIPAAm alone or by free polymer solutions of DTAB at such low concentrations ($c_s$< 3 mM). The correlation between foam stability and surface tension is also clear from figs. 3 and 9.

In figure 12, results of free drainage experiments are presented. In these experiments the initial liquid fraction for all foams was fixed to $\phi_l$=0.25 and the mean initial bubble radius, $R_B$, was about 70µm. The volume of the liquid drained was followed by direct observation with a CCD camera as a function of time. Note that for both systems the drainage is faster for T=45°C than for T=20°C, which was expected but, for $c_s$= 1.6 mM the drainage characteristic time (arrows in figure 12) is about 6 times larger for T= 20 °C than for T= 45°C; for $c_s$= 2.82 mM it is 3 times larger. We performed measurements of relative viscosity, $\eta_{solution}/\eta_{water}$, in Cop-L/surfactant mixtures as a function of DTAB concentration (see ESI) and we observed that the maximum change of viscosity occurs for free surfactant polymer solutions, for which the viscosity changes by a factor of 1.2 when changing the temperature from 45 to 25 °C. The effect of temperature on bulk viscosity is small and thus, it seems that it is not what controls the drainage velocity. One could think that changes in the size of the aggregates which take place when the temperature crosses the LSCT, inside the confined media given by liquid films, could explain the observed changes of the drainage dynamics. In that respect, we can estimate the size of the Plateau borders (liquid channels between adjacent bubbles)[64],

$$r_{PB} = \sqrt{\phi_l R_B^2} \qquad (9)$$

being $r_{BP}$ the Plateau border radius. For our foams, $R_B$ = 70 µm and $\phi_l$=0.25 for the initial stage of the free drainage process, thus, from eq. (9), $r_{PB}$= 35 µm, this is 35 times larger than the larger aggregate size (~1 µm). For the final stage of the drainage process $\phi_l$=0.01, $r_{PB}$ > 7 µm which is seven times the larger aggregate size (note that because of coarsening, $R_B$ will be larger than the initial value of 70 µm). Thus it seems more plausible that some other effect is responsible for the observed free drainage behaviour. The only possible explanation left is that there is a modification of surface rheology as T crosses the LCST. It seems plausible that surface viscosity changes are responsible for the changes in drainage velocity and foam stability, however this has to be corroborated experimentally. Finally we mention here that the temporal dependence of the light transmitted through the foam samples, as shown in figure 11, could be used to follow the coarsening dynamics[65,66], $R_B(t)$~$I(t)$. A systematic study of foam dynamics and of stability and its relation with surface rheology of these Cop-L/DTAB mixtures is currently under way and the obtained results will be object of a future article.

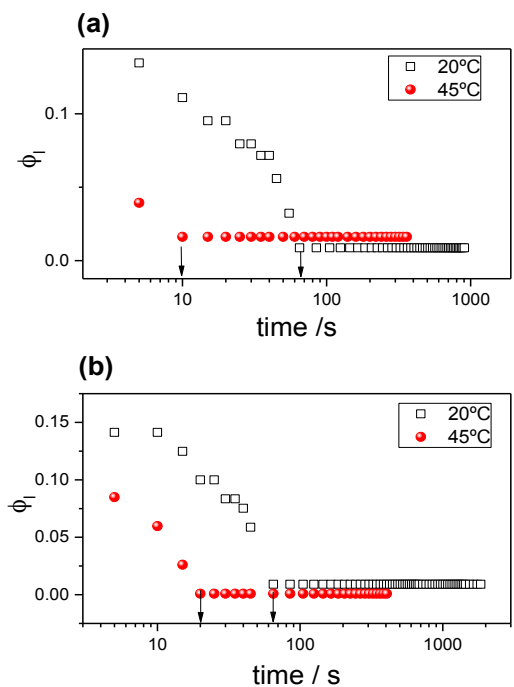

Figure 12 Liquid fraction as a function of time. (a) Cop-L 400 mg.L-1 + DTAB 1.62 mM; (b) Cop-L 400 mg.L-1 + DTAB 2.82 mM.

## 5. CONCLUSIONS

We studied a graft "co-polyelectrolyte" with a brush-type structure mixed with an oppositely charged surfactant as a function of surfactant concentration and temperature. By means of dynamic and static light scattering we found that, for $c_s<cmc$, the addition of DTAB, for T<LCST, produces a continuous collapse of the polymer chain, similar to what happens when the temperature is increased to values higher than the LSCT, in the absence of the surfactant. At concentrations over the *cmc* of DTAB, the aggregates increase their sizes instead of reducing them. We found that the aggregates formed are quite monodisperse although the polymer size distribution is somewhat broad, a fact that has been observed before[59].

From mobility and $\zeta$-potential measurements we constructed the binding isotherms and found that the aggregation process is non-cooperative up to $\beta\sim0.5$ and cooperative above said value. Because the sign of the $\zeta$-potential changes at very high surfactant concentration, we conclude that a fraction of the DTAB molecules bound to the polymer chain do so with their counterions and driven by hydrophobic interactions.

The measurements of equilibrium surface tension carried out on mixtures of Alg-PNIPAAm/DTAB in aqueous solutions demonstrated that the responsiveness of the copolymer to changes in temperature is preserved at liquid-air interfaces. An important point to be stressed is that this effect depends strongly on surfactant concentration. The relative change on foam stability is quite well correlated with the change in sizes of aggregates as measured by DLS, as well as with equilibrium surface tension changes as T crosses the LCST.

Despite not yet having results on interfacial dynamics (rheology), it seems that the effect of changing the temperature on the foam stability is due to changes on surface rheology. Despite not knowing without doubt the mechanisms involved, we demonstrated that the thermal responsiveness of the aggregates is conserved at the liquid-air interface and that those changes at interfaces have an effect on foams stability. The effect of surfactant concentration and temperature on the surface rheology of air/solution interfaces and its relation with foam stability, including free drainage and coarsening dynamics, is currently under investigation. The understanding of these complex systems in bulk is the first step in order to comprehend their behaviour at liquid-air interfaces and the properties of foams formulated with them. We think that these systems could be used for smart foams formulations, capable of responding to changes of temperature, if the correct surfactant concentration is chosen.

**Acknowledgements**

We thank Dominique Langevin for reading the manuscript and for valuable scientific discussions. This work was partially supported by grants PGI-UNS 24/F067 of Universidad Nacional del Sur and PICT 2013 (D) Nro 2070 of Agencia Nacional de Promoción Científica y Tecnológica (ANPCyT) and PIP-GI 2014 Nro 11220130100668CO (CONICET). CD and EFM thank CONICET for their fellowships.